# Unveiling Privacy and Security Gaps in Female Health Apps

MUHAMMAD HASSAN, University of Illinois Urbana Champaign, United States

MAHNOOR JAMEEL, University of Illinois at Urbana - Champaign, United States

TIAN WANG, Carnegie Mellon University, United States

MASOODA BASHIR, School of Information Sciences, University of Illinois at Urbana-Champaign, United States

Female Health Applications (FHA), a growing segment of *FemTech*, aim to provide affordable and accessible healthcare solutions for women globally. These applications gather and monitor health and reproductive data from millions of users. With ongoing debates on women's reproductive rights and privacy, it's crucial to assess how these apps protect users' privacy. In this paper, we undertake a security and data protection assessment of 45 popular FHAs. Our investigation uncovers harmful permissions, extensive collection of sensitive personal and medical data, and the presence of numerous third-party tracking libraries. Furthermore, our examination of their privacy policies reveals deviations from fundamental data privacy principles. These findings highlight a significant lack of privacy and security measures for *FemTech* apps, especially as women's reproductive rights face growing political challenges. The results and recommendations provide valuable insights for users, app developers, and policymakers, paving the way for better privacy and security in Female Health Applications.

## 1 Introduction

The recent Supreme Court decision in Dobbs v. Jackson Women's Health Organization, which effectively overturned Roe v. Wade, has intensified concerns regarding the privacy and autonomy of women's reproductive health in the United States [83, 96]. This ruling has resulted in the restriction or elimination of access to safe and legal abortion services in numerous states, raising alarms about the potential further erosion of privacy rights that were previously safeguarded under Roe. The implications of this legal shift extend beyond abortion access, affecting various dimensions of women's healthcare, including fertility treatments, contraception, and cancer care [78, 89].

The ruling has impacted the funding for female health clinics across some states in US [1, 47, 82]. Access to essential female health services has become increasingly challenging, often requiring individuals to travel long distances or

Authors' Contact Information: Muhammad Hassan, University of Illinois Urbana Champaign, Champaign, Illinois, United States, mhassa42@illinois.edu; Mahnoor Jameel, University of Illinois at Urbana - Champaign, Champaign, Illinois, United States, mjameel2@illinois.edu; Tian Wang, Carnegie Mellon University, Pittsburgh, Pennsylvania, United States, tianwan2@andrew.cmu.edu; Masooda Bashir, School of Information Sciences, University of Illinois at Urbana-Champaign, Champaign, Illinois, United States, mnb@illinois.edu.

even out of state [28]. Furthermore, individuals seeking these services may face harassment or legal repercussions [5]. Those who are financially disadvantaged, particularly low-income individuals and people of color, encounter even greater obstacles in accessing these vital services, leading many to seek alternative technological solutions in the form of *FemTech*[35, 52].

*FemTech*, a term referring to the application of technology to women's health and wellness, includes tools like Flo Health, Maven Clinic and Progyny. These technologies, designed to empower women by offering tailored solutions for managing reproductive health, have become central to many their lives [15, 61, 104]. These mobile applications, provide key services, including, such as period tracking, fertility and contraception support, menopause management, and sexual wellness education, but their deep integration into users' personal lives necessitates rigorous scrutiny of their privacy and data practices. As female health applications (FHAs) gain popularity, they bring with them a set of privacy and security challenges that could potentially undermine their intended benefits. In a world increasingly prone to data breaches and adversarial threats, the personal information stored by these apps — including menstrual cycles and fertility data — is especially vulnerable. This could jeopardize the very autonomy these tools seek to support. This introduces serious concerns, particularly given the heightened legal and social repercussions for reproductive health decisions in the post-Dobbs era.

In a digital landscape characterized by frequent data breaches and sophisticated adversarial attacks, the potential consequences of inadequate data security are severe. Adversaries who gain access to this data can infer intimate details about a user's reproductive health, such as their menstrual patterns, pregnancy status, or use of contraception. This information can be weaponized in numerous ways, from invasive surveillance and unauthorized data sharing to legal and social repercussions. The potential misuse of personal health data in this post-Dobbs landscape is particularly alarming. For instance, there have been documented case of women facing legal action based on their digital footprints, such as online purchases, text messages, or search histories related to abortion or miscarriage [10, 25, 100]. Such data leakage not only compromises personal privacy but also exacerbate existing inequalities, particularly for those who rely on FHAs due to limited access to traditional healthcare services. Consequently, some experts are even advising women to reconsider their use of FemTech apps for tracking reproductive health, given the risk that personal data could be used against them by authorities or under restrictive laws [10].

Female health applications (FHAs), while offering essential services like period tracking and pregnancy monitoring, collect sensitive health data that could expose users to significant privacy and security risks. Previous research has revealed that some FHAs have inadequate data security and privacy practices, including tracking and profiling users, sharing data with third parties, and failing to clearly disclose data practices to users [7, 58, 59, 80]. In certain jurisdictions, where women's reproductive rights are under intense scrutiny, this data could be exploited by authorities or other entities, as evidenced by cases like the Missouri state health department's tracking of women's menstrual cycles to determine whether abortions had taken place [55]. Given these concerns, it is essential to critically evaluate the security, privacy, and data practices of FHAs to understand their impact on users.

Prior research into the privacy and security aspects of FHAs has largely focused on a limited subset of applications, particularly those related to period tracking and fertility [7, 59, 80]. While these studies have been valuable, they do not capture the full complexity and scope of the FHA ecosystem. This paper addresses this gap by conducting a comprehensive analysis of the security practices, privacy measures, and data collection strategies employed by FHAs. We examine the specifics of the data collected by these applications and assess the level of transparency provided to users regarding these practices. By focusing on data privacy, we seek to illuminate the risks posed by these vulnerabilities and advocate for stronger protections in the FemTech sector.

This paper critically examines these escalating risks within the unique context of FHAs. It highlights how the nature of the data collected poses distinct threats, considering both individual and systemic impacts, particularly in marginalized and underserved lower-socioeconomic communities[20, 54]. Recent discussions, including those influenced by shifting legal and political landscapes focusing on the deep division on reproductive rights, underscore the urgency of addressing these vulnerabilities [43, 74, 96]. As policies evolve and public discourse intensifies, understanding the unique threats posed by these applications becomes even more crucial. Our study aims to bridge gaps in existing research by providing a comprehensive analysis of FHAs' data practices, security measures, and privacy policies. We will explore specific threat models relevant to this data and assess how breaches and misuse could affect users. This analysis will contribute to a more nuanced understanding of the risks and help advocate for stronger protections in the FemTech sector.

Given these escalating concerns, our study seeks to answer three critical questions that will illuminate the extent of these vulnerabilities and provide a roadmap for enhancing the security and privacy of female health applications.

- *RQ1: How do FHAs manage app permissions, and what privacy risks are associated with these permissions?*
  We aim to evaluate how FHAs handle app permissions and whether these permissions align with the app's functionality, while identifying associated privacy risks.
- *RQ2: What are the characteristics and purposes of third-party trackers embedded in FHAs, and what implications do they have for user privacy?*
  We will examine the types and functions of third-party trackers embedded in FHAs, such as analytics and advertising trackers, and assess their impact on user privacy.
- *RQ3: What types of user data are collected through dynamic interactions with FHAs, and how do these practices compare to the claims made in their privacy policies?*
  We intend to investigate the types of data collected during user interactions with FHAs and compare these practices to the disclosures made in their privacy policies.
- *RQ4: How accessible and comprehensible are the privacy policies of FHAs, and how well do they convey user rights and data management practices?*
  We plan to assess the readability and transparency of FHAs' privacy policies, evaluating their effectiveness in communicating user rights and data management practices.

## 2  Background

Technology for female health, commonly referred to as *FemTech*, encompasses the collection and processing of healthcare data specific to women, presenting both significant advantages and inherent risks. In this section, we conduct a comprehensive review of the existing literature to delineate the current landscape and identify the primary challenges facing *FemTech* in relation to data privacy and security. The section, explore the legal and policy frameworks that govern healthcare data in general, alongside those specifically related to female healthcare data, highlighting key factors that influence this domain. Furthermore, we assess the extent to which these frameworks adequately address the unique needs and concerns associated with the privacy of female healthcare data. Our objective is to uncover the existing gaps and potential opportunities for advancing data privacy and security in *FemTech*, informed by the prevailing literature and regulatory context.

### 2.1 Overview of *FemTech*

*FemTech* is a term that emerged in 2016 [4], encapsulating the application of technology to address female health concerns. This burgeoning field is dedicated to tackling a myriad of health issues that uniquely or disproportionately affect women, including maternal health, menstrual health, pelvic and reproductive health, fertility, menopause, contraception, and various other conditions [72, 102]. By leveraging technology, *FemTech* aims to empower women through personalized and accessible healthcare solutions, thereby enhancing their quality of life and reproductive autonomy.

The *FemTech* industry is experiencing rapid growth, with projections estimating a market value of $75 billion by 2025 [32]. A prominent category within this sector is female health applications (FHA), which are smartphone applications designed to provide a comprehensive suite of health services. These services include tracking, monitoring, and predicting health needs, as well as educating and advising users on various aspects of female health. FHAs can play a crucial role in supporting women with neonatal care during and after pregnancy [95], fertility and family planning [13, 30], menopause management, and other health-related needs. To deliver these services effectively, FHAs collect and process sensitive user information, encompassing menstrual cycles, fertility windows, pregnancy status, sexual activity, contraception use, hormonal levels, and various health conditions [53, 87, 92, 103].

### 2.2 Evolving Regulations and Political Pressures on Women's Healthcare in US

The U.S. healthcare system has long been shaped by complex historical and societal factors, resulting in unequal access to care for marginalized communities, particularly women of color. These disparities, rooted in structural inequalities across education, employment, and housing, have contributed to a well-documented mistrust in healthcare institutions among these groups [11, 90]. The escalating political rifts have further led to a gradual erosion of women's health rights and services in the U.S. Once-secure rights are now being systematically stripped away, with access to critical health services becoming increasingly obstructed. In cases where services are not outright eliminated, barriers are strategically erected to render them difficult, if not impossible, to access. The criminalization and infringement upon women's reproductive rights, rights that had been previously secured, represent a troubling reversal of progress [42, 71].

The increasing polarization of U.S. politics has undermined essential support for female health services. Federal and state rulings, often influenced by shifting governmental priorities and policy changes, have led to significant reductions in funding for these critical services. The politicization of female health has not only perpetuated bias and prejudice but has also disrupted efforts to achieve health equity for women, particularly those belonging to marginalized groups. As a result, women's health clinics across the country are facing severe financial constraints, with some state governments even resorting to legal actions against these facilities, accusing them of financial improprieties [1, 47, 82].

Women of color disproportionately bear the brunt of these regressive policies. The intersection of racial and gender disparities places women of color, especially those economically disadvantaged, at a heightened risk of adverse health outcomes. Research consistently shows that Black women and American Indian or Alaska Native women are 3.3 and 2.25 times more likely, respectively, to die from pregnancy-related causes than their white counterparts [90]. In an attempt to address these disparities, the FemTech industry has emerged as a technological solution to meet women's healthcare needs [44, 48]. However, despite its potential to fill gaps in healthcare access, FemTech, including female health apps (FHAs), has raised significant concerns about data privacy.

One of the critical issues with FemTech is its often-unregulated nature. Female health apps, which collect sensitive reproductive health data, are not covered by HIPAA (Health Insurance Portability and Accountability Act) regulations [75, 77]. This lack of oversight allows these apps to harvest, share, and, in some cases, exploit personal data without adequate accountability. Studies have shown that many of these apps engage in data practices that may compromise user privacy, contributing to an emerging trend of surveillance, or "dataveillance," of women's reproductive health data [46, 70]. These findings underscore the urgent need to address the privacy and security risks associated with FHAs and ensure that women's reproductive health data is protected from exploitation.

### 2.3 Data Exploitation and Security Risks in U.S. *FemTech*

In the United States, female health applications (FHAs) are at the forefront of a rapidly evolving digital health landscape, aggregating sensitive, personally identifiable health data, that unveils the intimate details of women's lives, preferences, and identities. Trust and data safety are paramount for users of these technologies. While fostering trust is essential for encouraging users to share their personal information [3, 45], this trust is often precariously built on a foundation of inadequate user awareness regarding the consent they provide and the full implications of data sharing [40, 93, 99].

Data privacy transcends the mere safeguarding of information; it is fundamentally intertwined with the safety, dignity, and autonomy of women. Users place immense trust in these applications by sharing sensitive details about their sexual and reproductive health (SRH). Trust is paramount in fostering inclusivity and mitigating privacy risks for marginalized communities, including women of color and those facing financial hardships, who often experience the significant repercussions when access to female reproductive health facilities is curtailed or restricted[94, 97].

The implications of misusing or exposing this confidential information are severe and far-reaching. The risks are not merely theoretical; "dataveillance" of FHAs can lead to discrimination, harassment, stigma, violence against women, and even jailtime in states, based on their SRH status or choices. Such outcomes highlight the urgent need for user-centered privacy and data protection measures that prioritize the rights and well-being of users. As the landscape of reproductive health continues to evolve, it is imperative that we recognize the profound impact of data privacy on the lives of women, particularly those from vulnerable communities. The stakes are high, and the call for action is clear: safeguarding women's health data is not just a technical challenge; it is a moral imperative that demands immediate attention and action [8, 53, 65].

The challenges and risks associated with FemTech and FHAs in the US have been well-documented, particularly the troubling reality that these technologies often collect and process sensitive information without adequate consent or protective measures [8]. Many users remain unaware of the complex trade-offs involved in granular consent models and secondary data usage, including how their data will be used, shared, and potentially exploited, as well as the associated benefits and risks [40, 93, 99].

The misuse of sensitive FHA information carries severe legal and social consequences. For instance, in 2019, the state of Missouri sought to obtain medical records from patients who had undergone abortions at a Planned Parenthood clinic in St. Louis, as part of an investigation into alleged legal violations [55]. This invasive action highlights the precarious state of data sovereignty over women's health data, particularly in the wake of the Supreme Court's decision in *Dobbs v. Jackson Women's Health Organization*, which overturned *Roe v. Wade* and allowed states to implement stringent abortion bans [17, 64]. In this new legal landscape, the algorithmic governance of FHA data has become critically important, as the ways in which reproductive health data is collected, processed, and shared can have direct legal and social consequences. The exposure or misuse of this data by state authorities or anti-abortion activists not only risks legal repercussions but also has a chilling effect on user trust in digital health technologies. As a result, many women may hesitate to engage with FHAs out of fear of privacy breaches or being criminalized for their reproductive health choices, further infringing on women's reproductive rights and autonomy [57, 73]. This erosion of trust threatens the potential for technology to play a supportive role in women's health, underscoring the need for stronger privacy by design approaches to protect users' rights and data.

The commodification of data collected by FHAs—including menstrual cycles, fertility, pregnancy, and sexual health—introduces significant risks of unwanted profiling, discrimination, and exploitation. This information can be used by advertisers, marketers, researchers, and insurers to target and manipulate users based on their reproductive health (SRH) needs or interests. For instance, a recent investigation by Consumer Reports revealed that certain FHAs shared users' reproductive data with third parties, violating promises of confidentiality [63, 76]. Furthermore, the Federal Trade Commission (FTC) uncovered instances where apps disclosed users' health information to platforms like Facebook and Google without obtaining explicit user consent [21]. These breaches of trust enable harmful real-world consequences such as economic discrimination, digital surveillance, and biased health outcomes, as users are subjected to practices that can lead to stigmatization or denial of services based on their SRH data.

This study focuses on female health apps (FHAs) within the United States, a unique case study due to North America's dominant role in the FemTech market. The U.S. hosts the majority of *FemTech* companies and receives the largest share of investment in the sector, making it a critical environment for examining both the challenges and opportunities in this space [85, 86]. A U.S.-centric analysis enables a deep dive into the regulatory disparities in FemTech governance, as well as the specific privacy and security concerns that emerge in a landscape shaped by recent legal and political shifts. Although our primary analysis focuses on U.S. FHAs, the subsequent subsection will briefly explore geopolitical variances in digital health innovation and global trends in *FemTech*, providing context for our findings while maintaining a clear focus on the U.S. market.

### 2.4   Global Context of Female Health Data Privacy

The global landscape of female health data privacy demonstrates regulatory heterogeneity. Key international frameworks, such as the GDPR, significantly shape how data privacy is managed across regions. This study focuses on the U.S. context, given North America's significant stake in the FemTech market, aiming to analyze the specific data privacy risks and governance structures for U.S. users, while global trends inform our comparative understanding.

In recent years, global female health rights and services have evolved, though systemic health access disparities persist [14, 38]. Collaborative governance between organizations such as the United Nations Population Fund and Médecins Sans Frontières, alongside governmental bodies, has led to digital health equity initiatives aimed at enhancing reproductive, maternal, and adolescent health services. Advancements in data protection and privacy-preserving health governance through global health data sovereignty frameworks have increasingly focused on safeguarding sensitive female health information. The GDPR in the European Union set a global standard for privacy-preserving health governance [101], with South Africa's POPIA and other countries like India(DISHA act), UAE(Health Data Law), and Nigeria following suit with similar regulations.

In contrast to global progress, the U.S. faces regulatory inertia, despite its leading role in FemTech services. While it boasts the highest number of FemTech companies worldwide, many of these firms engage in data surveillance and commodification practices, sharing sensitive user data with third parties, raising critical concerns about the security of female health data [84]. This growing disparity between U.S. practices and global transnational privacy frameworks highlights an urgent need for systematic investigation into the security vulnerabilities and data handling practices of female health apps. Addressing these risks is essential for ensuring the safety and privacy of users within female healthcare technology.

## 3 Methodology

Our research focuses on understanding the privacy and security risks posed by female health applications (FHA). We aim addresses the following objectives: (1) conduct static analysis of applications to assess potential capabilities, data access, and risks; (2) perform qualitative and linguistic analysis of privacy policies to evaluate their content, readability, and alignment with user rights; and (3) capture dynamic interactions with the applications to identify the types of data collected from users and the sources of such data. These methodologies allow us to investigate the often adversarial relationship between user privacy and app functionality in female health apps, a domain of increasing importance in HCI and privacy research [16, 50, 51].

To answer our research questions, we first selected a representative sample of FHA from the Google Play Store, following a simulated user search strategy. We then conducted static analysis to examine app permissions, uncovering the technical behaviors that may pose privacy risks, followed by an exploration of third-party trackers embedded within the apps. In parallel, we manually interacted with each app to document the types of user data requested and collected, such as personal identifiers, health data, and location information. Lastly, we analyzed the privacy policies of the apps, focusing on their scope, accessibility, and compliance with best practices for transparency and user rights.

This mixed-method approach, combining quantitative data-driven insights with qualitative privacy policy evaluation, provides a comprehensive understanding of the privacy risks users face when engaging with FHA. Our methodology enables us to bridge the gap between technical app behavior and the user-facing aspects of privacy, offering novel insights into the under-researched domain of female health apps in the context of privacy and data security.

### 3.1 App Selection

We selected Android-based FHA from the Google Play Store due to its dominant market share ( 70%) [79]. Since the Play Store lacks a dedicated category for female health, we simulated real user behavior by searching terms such as *"woman"*, *"female"*, and *"feminine"*, in combination with *"health"*, *"wellness"*, and *"well-being"*. Applications were filtered with US as geo-location and filtered based on download count and rating. To refine our dataset, we applied the following criteria:

- At least 5K downloads,
- Explicitly targeted at female users (based on app name/description),
- Free to download.

We collected a sample apps in June and again in October 2023 and then in March 2024, ensuring a temporal snapshot of female health apps. After filtering out duplicates, our final dataset consisted of 45 apps. Our selection of appls follows the research methodology of previous studies, such as Adhikari et al. and Shipp et al. [2, 80]. By adopting both a technical analysis of permissions and trackers, and a qualitative examination of privacy policies, we provide a more comprehensive view of the risks these apps pose. Additionally, our sample size exceeds those used in prior research, enhancing the robustness of our findings.

### 3.2 Static Analysis of Permissions

Permissions are critical for app functionality but can pose significant privacy risks when mismanaged [31, 49]. To investigate how FHA apps handle permissions, we conducted static analysis using the Android Debug Bridge (ADB) to extract APK files, which were then analyzed with Androguard, a reverse-engineering tool for Android applications [36]. Our analysis categorized the requested permissions according to Google's official classification [9]. The categories include Normal Permissions (low-risk, minimal privacy impact), Dangerous Permissions (high-risk, access to sensitive data), Signature Permissions (granted when apps share a signing certificate), and SignatureOrSystem Permissions (for system apps or shared vendor services). Detailed descriptions are provided in Table 1.

| Permission Type | Description |
|---|---|
| Normal Permissions | Low-risk permissions that have minimal impact on user privacy or system integrity. |
| Dangerous Permissions | High-risk permissions that grant access to sensitive data, such as personal and health-related information. |
| Signature Permissions | Automatically granted if the app shares the same signing certificate with another app. |
| SignatureOrSystem Permissions | Intended for system apps or shared vendor services. |

Table 1. Categorization of Requested Permissions Based on Google's Official Classification

### 3.3 Third-party Trackers and Libraries

The integration of third-party libraries is a common practice in FHA apps, often used for purposes such as analytics, advertising, and monetization. However, the use of third-party trackers can significantly increase privacy risks in FHA apps, as these trackers often operate outside the control of the app developers. Trackers may collect sensitive health and behavioral data, leading to potential privacy violations, especially when they are integrated for monetization and behavioral profiling [80]. By embedding trackers that gather user data for external parties, FHA apps can inadvertently facilitate data leakage or unauthorized data sharing. Therefore, these libraries can introduce privacy risks by collecting user data without explicit user consent [12]. To assess the presence and purpose of such trackers, we used Exodus Privacy [29], a tool that detects embedded third-party trackers in app code. Trackers were categorized based on their functionality, such as advertising or analytics. For those not categorized by Exodus, we performed manual verification through documentation provided by the tracker providers.

### 3.4 Dynamic Interaction and User Data Collection

To comprehensively assess the scope of user data collection, we performed manual dynamic interactions with each app, simulating a typical user journey. Our goal was to capture all data requests made by the apps, from basic personal identifiers (e.g., name, email) to highly sensitive health-related data. This manual approach allows for a nuanced and complete understanding of the user experience, which automated tools alone may miss.

Each app was explored for 10-15 minutes or until all available features for a free user had been exhausted. We documented the types of data requested across various app screens, including sign-up flows and in-app questionnaires, and classified the data based on its sensitivity (e.g., personal identifiers, health data) and its intended purpose. Furthermore, we recorded the features offered by the apps to understand the context in which this data was being collected.

Given the political and social implications of reproductive health data, this analysis is critical to understanding potential risks to users, such as discrimination, harassment, or prosecution [10, 25, 78, 89, 100]. The collection of this sensitive data must be considered in relation to each app's technical attributes (e.g., permissions, trackers) and compliance with privacy regulations (e.g., privacy policies).

To ensure the accuracy and reliability of our findings, the interactions were carried out independently by three reviewers, and any difference was resolved at the end. This multi-reviewer approach mitigates bias and allows for a more holistic capture of data collection practices across different use cases. The combined manual exploration and independent review process enhances the depth and completeness of our dataset.

### 3.5 Privacy Policy Analysis

Privacy policies serve as critical documents for informing users about the data practices of apps. However, these policies are often complex and difficult to interpret, creating barriers for users to fully understand how their personal data is handled [37, 67]. To assess the transparency and clarity of privacy policies in FHA, we evaluated them based on accessibility, user rights, and readability, aligning with best practices for privacy communication in HCI and data ethics.

*3.5.1 User Rights and Transparency.* We employed the Fair Information Practice Principles (FIPPs) as the guiding framework to assess how user rights are articulated within the privacy policies [33]. FIPPs, unlike HIPAA or GDPR, provides a flexible and widely applicable set of principles for privacy best practices. HIPAA specifically governs healthcare entities, and at the time of writing, is not applicable to FHA. Similarly, GDPR, while offering robust privacy protections, primarily targets EU-based services, whereas the scope of this study is FHAs in U.S. context. Hence, FIPPs offers a more relevant framework for evaluating privacy practices across a broad spectrum of services, including FHA.

We evaluated user rights such as access, correction, deletion, portability, and consent, all of which are fundamental to ensuring user agency over personal data. Furthermore, we examined whether the policies include additional rights, such as the ability to opt out of data collection or to file complaints, extending beyond the standard FIPPs framework. This comprehensive analysis provides insights into the transparency and accountability of FHA developers with regard to user privacy.

*3.5.2 Policy Accessibility and Readability.* The accessibility of privacy policies is essential for users to make informed decisions. We first assessed where these policies are located—either within the Google Play Store's Data Safety section or on the developer's website. This step highlights how easily users can access important privacy information prior to app installation.

To further analyze the usability of these privacy policies, we assessed their readability using the Flesch-Kincaid Reading Ease and Grade Level scores. These metrics quantify how difficult a text is to read based on sentence and word complexity. A lower Flesch-Kincaid Reading Ease score suggests more complex text, while the Grade Level score corresponds to the educational level required for comprehension. In the U.S., the average reading level is at an 6th-grade level, and texts requiring a 12th-grade reading level or higher are considered difficult for the general population to understand [22, 41]. By comparing these scores, we can evaluate whether FHA policies are written in a manner accessible to a wide audience or if they pose unnecessary barriers to comprehension.

Through these analyses, we aim to determine whether FHA privacy policies offer a clear, user-friendly, and legally compliant view of data practices, and how effectively they communicate user rights and data management practices to their target audience.

## 4 Results

In this section, we provide an overview of the 45 female health applications (FHAs) analyzed in this study, encompassing over 587 million downloads from the Google Play Store. Notably, two of these applications surpass 100 million downloads, with the average number of installs across the dataset exceeding 13 million. Fig 1 presents the distribution of download ranges for all FHAs included in the study.

| Category | Unique Permissions | # FHA |
|---|---|---|
| **Normal** | 30 | 45 |
| **Dangerous** | 27 | 42 |
| **Unknown** | 58 | 44 |
| **Signature** | 3 | 40 |

Table 2. Permission Restriction Categories and FHA

| Functionality | # FHAs |
|---|---|
| Fertility/menstruation/pregnancy | 35 |
| Diseases/illnesses screening and symptom logging | 10 |
| Mental health (self-care) | 3 |
| Exercise | 16 |

Table 3. App Functionality in Female Health Apps

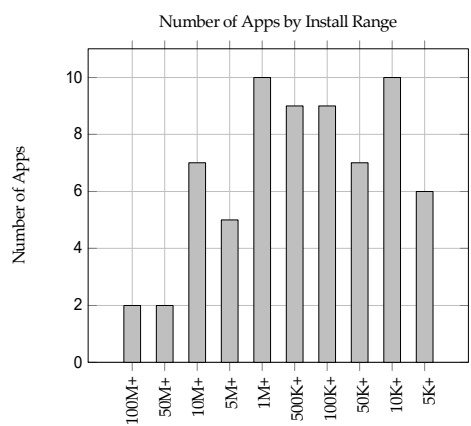

Fig. 1.

Bar chart showing the distribution of apps according to their install ranges. The x-axis represents different install ranges from 100 million to over 500,000 installs. The y-axis shows the number of apps, ranging from 0 to 10. Each bar corresponds to an install range category and indicates how many apps fall into that category.

The analysis also reveals the presence of a developer with two high-download applications, collectively accounting for more than 115 million downloads. These two applications, which rank 3rd and 5th in our dataset by number of downloads, as shown in 4, are categorized under the *Health and Fitness* section of the Google Play Store.

A majority of the apps (84%) are categorized under *Health and Fitness,* with a smaller proportion (15%) labeled as *Medical* and one app classified under *Lifestyle.* To facilitate a more nuanced understanding of their functional claims, we manually assign additional labels based on the services described by the applications. This granularity allows us to more accurately capture their core offerings, as some apps claim multiple functionalities.

Table 3 outlines the primary functionalities observed across the dataset. Unsurprisingly, fertility, menstruation, and pregnancy-related services emerge as the most prevalent, with 78% of apps offering at least one of these features. Furthermore, 22% of apps claim to track diseases and symptoms, such as breast cancer and reproductive organ health, while 6% provide mental health support tailored specifically to women. Additionally, 35% of the apps offer fitness services, including pelvic floor exercises, yoga, strength training, and weight tracking, often combined with their core functionalities.

For brevity in reporting the results and discussions, we refer to each application by its corresponding app ID (#) from Table 4 in the remainder of the paper. This allows us to maintain clarity while discussing individual app behaviors and characteristics.

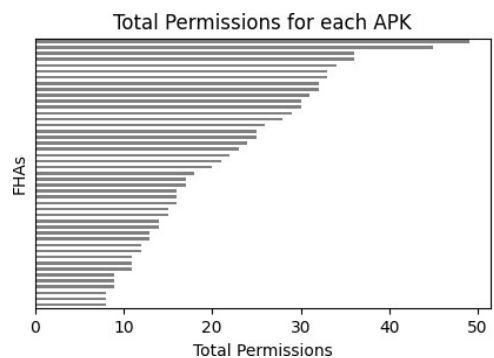

Fig. 2. Total Permission Distribution across FHAs

## 4.1 Permissions Analysis

Permissions define the level of access an application has to user data and device functionality. In Android, apps can request various permissions to function, but previous research shows that many apps often request more permissions than necessary, raising privacy concerns [26]. These concerns are heightened by the potential misuse of personal data, lack of transparency, and the growing fear of surveillance [18, 23].

*4.1.1 Distribution of App Permissions.* Every FHA in our dataset utilizes one or more *Normal* permissions, with 28 unique *Normal* permissions identified. The most privileged app in terms of the number of permissions is #17, *Newfemme*, which requests 49 permissions and has approximately 1.34 million downloads with a 4.3 rating on Google Play. The second-highest is app #2 with 45 permissions, exceeding 1.39 million downloads and a 4.3 rating. Apps #33 and #37 follow closely, each requesting 36 permissions, though their downloads are significantly lower, at 51,000+ and 19,000+, respectively.

In contrast, the least privileged apps (#14, #42, and #45) request only 8 permissions each, with downloads ranging from 10,000 to 100,000+. All apps request more permissions than the average Android app, with an average of 21 permissions per app in our dataset (standard deviation $\approx 10$), as the permission distribution can be seen in fig 2. There is considerable variation in the permission privileges of FHAs, even though many of them provide similar services, as illustrated in Table 3. This underscores significant inconsistencies in permission usage among FHAs. Furthermore, every application in our dataset requests eight or more permissions, exceeding the reported average of five permissions for Android apps [66, 68].

The most frequently requested permissions include *WAKE_LOCK* (45), *INTERNET* (45), *ACCESS_NETWORK_STATE* (44), *BIND_GET_INSTALL_REFERRER_SERVICE* (41), and *RECEIVE* (40). Conversely, 41 permissions are used only once, including health-specific permissions such as *READ_BLOOD_PRESSURE, READ_RESPIRATORY_RATE, READ_SEXUAL_ACTIVITY, READ_BODY_FAT, READ_CERVICAL_MUCUS* and others related to body metrics. These outlier permissions raise questions about the appropriateness and necessity of certain data access requests in FHAs.

A detailed breakdown of permissions is presented in Appendix 7.

*4.1.2 **Permissions and Their Categorization**.* Permissions in Android apps are grouped based on the type of data they access, ranging from normal, signature, to dangerous and unknown permissions. In our analysis of the 45 FHAs, we identify 118 unique permissions. These permissions are categorized as *Normal*, *Signature*, *Dangerous*, and *Unknown*. The *Normal* permissions pose minimal risk, while *Dangerous* permissions have the potential to expose sensitive user information [9]. Notably, we also encounter a subset of permissions labeled as *Unknown*, which cannot be definitively categorized based on available Google Developer documentation, indicating potential gaps in transparency from app developers. Our analysis shows that the use of permissions varies widely across FHAs, with many apps requesting access to sensitive data, while others show more restrained usage.

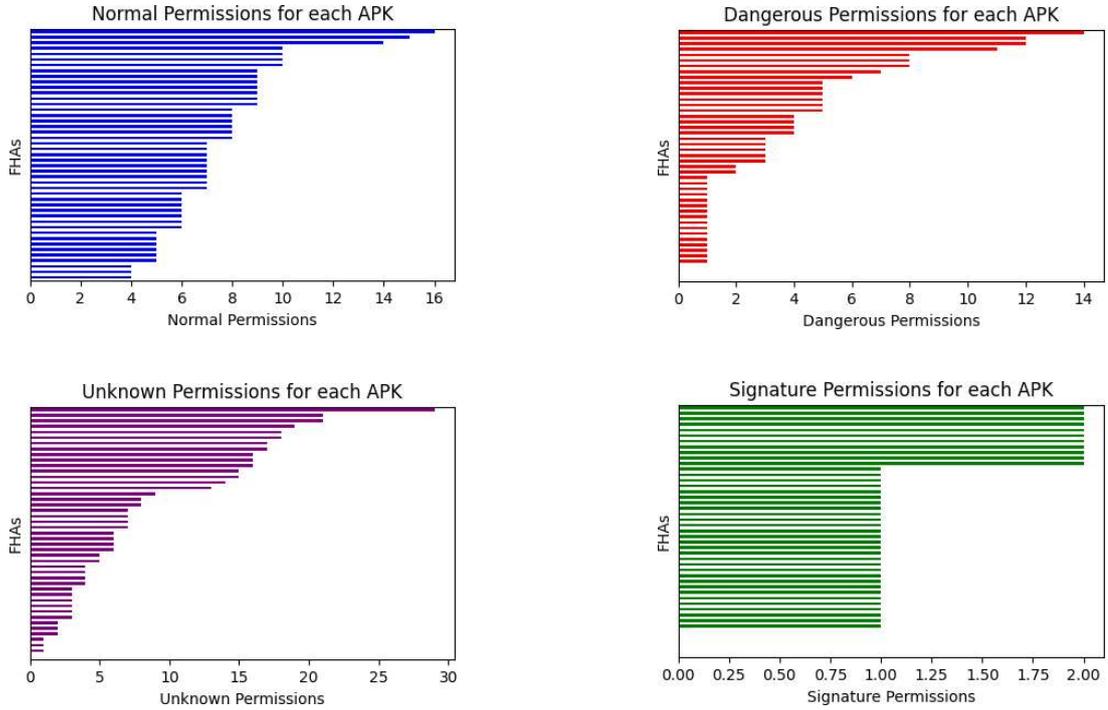

Fig. 3. Permission Distribution based on categorized identified.

**Normal Permissions** are the most frequently observed across our dataset, appearing in all 45 FHAs. Commonly requested permissions in this category include *WAKE_LOCK*, *INTERNET*, *ACCESS_NETWORK_STATE*, *RECEIVE_BOOT_COMPLETED*, and *FOREGROUND_SERVICE*. These permissions are generally considered low-risk, as they enable basic app functionality without exposing critical user data.

In contrast, **Dangerous Permissions** are found in 42 apps, raising privacy concerns due to the sensitive nature of the data they provide access to. The most commonly requested dangerous permissions are *POST_NOTIFICATIONS*, *WRITE_SETTINGS*, *READ_EXTERNAL_STORAGE*, *WRITE_EXTERNAL_STORAGE*, and *READ_MEDIA_IMAGES*. While these permissions are often justified for core functionalities, such as sending notifications or accessing media files, there are instances where permissions like *READ_CONTACTS*, *READ_PHONE_STATE*, and *CALL_PHONE* are requested without clear, legitimate use cases. For example, FHAs #11, #30, #33, and #34 include such permissions, potentially introducing privacy risks by accessing contact lists or initiating phone calls without user awareness. The FHAs #11,

#17, and #35 stand out for requesting the highest number of dangerous permissions, correlating with their high overall permission counts.

The **Unknown Permissions** category, comprising permissions that cannot be definitively classified using Google's developer documentation, is also prevalent, appearing in 44 apps. Many of these permissions pertain to system-level functionalities, such as managing advertising IDs, setting notification badges, or accessing biometric data. While these permissions might serve necessary functions, their ambiguity poses additional risks. Notably, FHAs #2 and #38 include health-specific unknown permissions, which could expose highly sensitive user health data if mismanaged (see Table 5). Among the dataset, apps like #29, #40, and #14 do not request any dangerous permissions, demonstrating a more privacy-conscious design. Additionally, app #45 does not utilize any unknown permissions. We also identify three unique *Signature Permissions*, which are present across 40 apps, often providing access to specialized app-level functionality.

Table 2 summarizes the distribution of unique permissions across the Normal, Dangerous, and Unknown categories, highlighting the number of FHAs requesting permissions in each category and figure 8 shows permission across all FHAs.

### 4.2 Prevalence and Categorization of Third-Party Trackers

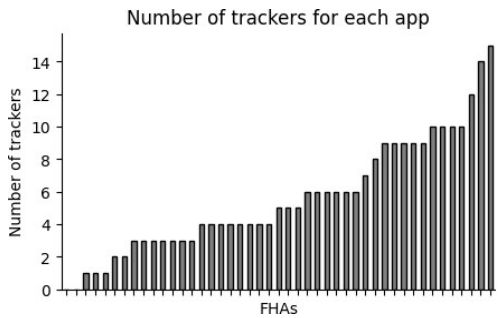

Fig. 4. Total Number of Trackers in FHAs

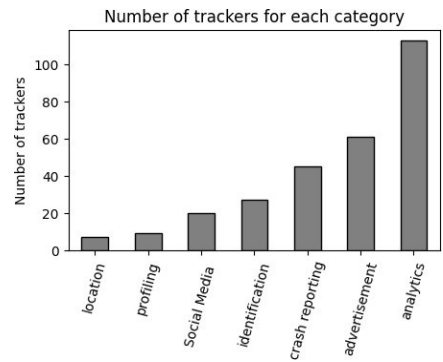

Fig. 5. Frequency of Trackers in each Category

Our analysis reveals the widespread presence of third-party trackers within FHAs, with a total of 252 occurrences across the dataset and 51 unique trackers identified. Trackers, which collect and transmit user data, are embedded in 95% (43 out of 45) of the FHAs in our sample. The number of trackers varies, ranging from 1 to 15 per app, with an average of 7 trackers per FHA. This extensive tracker usage highlights significant privacy concerns, as illustrated in fig 4, which outlines the variance in tracker distribution across FHAs.

The trackers are categorized based on their primary function, with many serving multiple purposes. These categories include *Analytics*, *Advertisement*, *Crash Reporting*, *Identification*, *Social Media*, *Profiling*, and *Location*. Fig 5 provides a breakdown of the number of trackers and their frequency in each category.

*4.2.1 Distribution of Trackers Across FHAs.* The FHA with the most extensive use of trackers is app #15, embedding 15 unique trackers. With over 2 million downloads, this app's trackers span five categories, including 5 for *Analytics*, 10 for *Advertisement*, 2 for *Crash Reporting*, 2 for *Identification*, and 1 for *Profiling*. Similarly, FHA #6 includes 14 trackers across similar categories, while app #33 incorporates 12 trackers from all major categories. Notably, apps #39 and #41 are the only FHAs in the dataset without any embedded trackers, underscoring the variance in privacy practices across the ecosystem. The figure 8 shows trackers across all FHAs

*4.2.2 Dominance of Trackers from Major Tech Companies.* A striking finding in our analysis is the dominance of trackers from a small number of major technology companies. The most prevalent tracker is *Google Firebase Analytics*, embedded in 41 of the 45 apps, falling under the *Analytics* category. *Advertisement* trackers are the second most common, followed by those for *Crash Reporting*, *Identification*, *Social Media*, and *Profiling*.

Google and Facebook are by far the most dominant entities, accounting for 64% of all tracker occurrences in our dataset. Specifically, we detect six different types of Facebook trackers, representing 72 occurrences (29% of all trackers). Google has five unique trackers, appearing 88 times (35% of all occurrences). The overwhelming presence of trackers from these two companies suggests that they collect vast amounts of sensitive user information, enabling them to build detailed user profiles for targeted advertising and other purposes. This monopolization of user data not only raises privacy risks but also concentrates significant power in the hands of a few corporations.

While Amazon and Microsoft trackers are present in our dataset, their footprint is considerably smaller. We identify three distinct Amazon trackers across 5 FHAs and two Microsoft trackers in 3 FHAs, indicating a more limited role for these companies in the female health app ecosystem.

*4.2.3 Privacy Implications of Tracker Prevalence.* The extensive use of third-party trackers, particularly those from dominant platforms like Google and Facebook, poses significant privacy risks. These trackers collect sensitive user data, which can be used to create detailed profiles of individual users, potentially infringing upon their privacy. Moreover, the concentration of data collection by a few large companies raises concerns about monopolization and the lack of diversity in data control. Users of FHAs are often unaware of the extent to which their data is being shared, and the high presence of these trackers raises questions about informed consent and the transparency of these practices. The ramifications for women's health data, which is particularly sensitive, are profound, highlighting the urgency for better privacy protections and regulatory scrutiny.

### 4.3 Privacy Policy Analysis

Privacy policies are intended to inform users about how applications collect, store, and share their data, along with the rights and choices available to users concerning their personal information. However, many FHAs fall short in both the accessibility and the transparency of their privacy policies. To evaluate this, we analyze the privacy policies of the FHAs in our dataset across four key dimensions: *availability*, *scope*, *readability*, and *data practices*, section A in appendix lists detailed factors investigated for each policy. This analysis sheds light on the gap between the stated privacy practices and the actual behavior of these applications, which poses significant concerns regarding user privacy and data protection.

*4.3.1 Availability of Privacy Policies.* Ideally, privacy policies should be easily accessible to users before they decide to install an application, giving them the opportunity to make informed decisions about their data privacy. For Android applications, this often means having a privacy policy clearly displayed on the Google Play Store page.

Upon investigating the availability of privacy policies for the FHAs in our dataset, we find that 15% of these apps do not provide a privacy policy on the Google Play Store. These applications account for a combined total of more than 14.6 million installs, leaving a significant portion of users with little to no information regarding how their data is managed. For these apps, we seek privacy policies on their respective websites but could not find any policy for 8% of FHAs. This lack of transparency in the initial stages of user engagement remains a critical issue, particularly for apps handling sensitive female health data.

*4.3.2 Scope of Privacy Policies.* Our analysis extends to the scope and specificity of the privacy policies available. Policies should ideally be dedicated to the specific app in question, addressing the unique features, data collection practices, and security measures of that app. Unfortunately, we discover that 60% of the FHAs use generic or non-specific policies that do not account for the unique functionalities of the app. Half of these are shared policies used across multiple apps or services by the same developer, while the other half are entirely generic documents primarily focused on the developer's website policy.

Furthermore, of the apps that have privacy policies available on the Google Play Store, only 55% have policies that are dedicated to the specific FHA under investigation. The remaining apps either use broad, vague language or avoid referencing the specific functionalities of the FHA, raising concerns about transparency and accountability.

*4.3.3 Readability of Privacy Policies.* A key aspect of privacy policy effectiveness is readability—users should be able to easily understand the terms and conditions regarding their data privacy. Best practices suggest that privacy policies should score at least 60 on the Flesch Reading Ease scale, equating to an 8th-9th grade reading level, to ensure accessibility to a broad audience.

However, our analysis finds that the privacy policies of FHAs are far from this benchmark. The highest recorded Flesch Reading Ease score in our dataset is 46.28, while the average is a mere 31.1, suggesting that these policies require a college-level education to fully comprehend. Additionally, the average policy length is approximately 4,000 words and 150 sentences, making them long and cumbersome to read. These findings are consistent with earlier research showing that many mobile health app privacy policies are excessively complex and challenging for the average user to understand [88]

The low readability of these privacy policies poses a substantial barrier to user comprehension, particularly for marginalized or less-educated populations, potentially exacerbating the risks these users face in making uninformed privacy decisions.

*4.3.4 Data Practices.* Our investigation into the data practices outlined in the privacy policies focuses on two primary areas: data sharing and user rights regarding data regulation and control.

- **Data Sharing:** An alarming 85% of the FHA privacy policies explicitly or implicitly acknowledge that they share user data with third parties. However, 80% of these policies provide little to no details on which third parties receive the data, the specific information being shared, or the purpose behind the data sharing. Only 15% of the policies clearly outline the responsibilities of third-party recipients or company employees who access user data, and fewer than 10% indicate any requirements for training in handling sensitive user data.

- **Data Regulations and User Rights:** While 90% of the FHAs claim that users have the right to access their data, only 50% of the policies specify how long they store user data, often leaving it ambiguously tied to business needs. Similarly, 85% of the policies mention users' rights to correct their data, but half of the apps do not clarify whether user consent is updated in the event of account deactivation or deletion. Regarding compliance with regional privacy regulations like GDPR or CCPA, 65% of the policies claim adherence, but only 6% describe how they process data for teenage users.

These findings illustrate a concerning disconnect between the privacy policies of FHAs and the protection of user data, with many policies providing inadequate information and insufficient assurances regarding data management.

### 4.4 User Data Collection and Dynamic Interaction Analysis

In this subsection, we analyze the types of user-generated information collected by Female Health Apps (FHAs) during both the installation and interaction phases. Such user-provided data is essential for these apps to function, given their focus on reproductive and health-related services. However, this data is highly sensitive, which raises concerns about privacy and data security. Our investigation categorizes the user information requested by these apps, as shown in Table 6. Two researchers independently label the data points, followed by a joint reconciliation process to ensure consistency, ultimately identifying five main categories of user information.

- **Demographic Information**: *Demographic information* is the most frequently requested category, with 43 FHAs requesting at least one type of demographic data. Common data points in this category include users' Name, Profile Picture, Email, and Age or Date of Birth. Additionally, 6 FHAs request users' Race/Ethnicity, while fewer apps (3) ask for employment details, and only 2 apps seek income information. One FHA even asks for users' experiences with discrimination or financial constraints. Given the sensitive nature of health and reproductive data, users may feel particularly uneasy about sharing their work history, financial details, or experiences with discrimination through these apps.
- **Physical Health Information**: *Physical health information* is the second most requested category, with 41 FHAs inquiring about users' physical health. Weight is the most requested data point (38 apps), followed by exercise activities (32 apps). Additionally, 31 apps ask for information on users' existing or past health conditions, and 23 apps inquire about any medications currently being taken. These health histories encompass serious conditions, such as gastrointestinal diseases, cancer, respiratory issues, and reproductive health concerns. Notably, 3 FHAs allow users to share and track family members' health statuses, and 13 apps integrate wellness shopping features within the app.
- **Reproductive Health Information**: *Reproductive health information* is the third most common category, with 39 FHAs requesting data. Pregnancy tracking is particularly prevalent, with 35 apps asking for this information. Similarly, 33 FHAs track menstrual cycles, while 31 apps inquire about users' sexual activity. Less frequently requested reproductive data includes Hormone Therapy information (10 apps), OB-GYN history (8 apps), and whether a user's partner uses contraceptives (3 apps).
- **Mental Health Information**: *Mental health information* is another significant category, with 36 FHAs requesting users' mental health data. Commonly requested data points include mental health conditions (33 apps), mood tracking (33 apps), stress level tracking (31 apps), and sleep information logging (29 apps).
- **Alcohol and Drug Activity Information**: In the *Alcohol and Drug Activity* category, 15 FHAs collect data on users' consumption habits, including cigarette smoking (12 apps) and travel patterns (7 apps).
- **Preventive Health Information**: Only 4 FHAs request *preventive health information*, specifically asking users to log vaccination records and dates.

*4.4.1 Core App Features.* In this section, we manually map the collected permissions to the core functionalities of the FHAs recorded during dynamic interaction with the apps. Two reviewers collaboratively label each permission as either being necessary for the core functionality of the app or classify it under 'other' if it does not align with core functions.

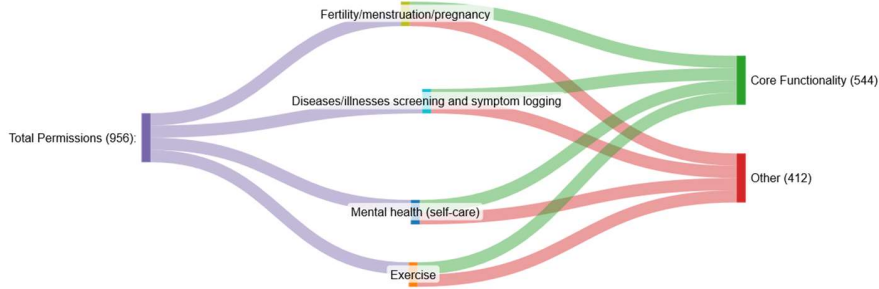

Fig. 6. Mapping of FHAs Permissions to Core Functionality

We analyze a total of 956 permissions across the FHAs for their claimed functionalities (functions as shown in Table 3). Of these, 544 permissions (57%) are deemed essential to the app's core functionality, as described in the app's Google Play Store listing or observed during app interactions. However, 412 permissions (43%) are classified as 'other,' indicating they are not transparently linked to the core functionality, as shown in 6. It is important to note that permissions labeled as 'other' do not necessarily indicate a privacy or security risk but highlight a lack of transparency regarding the app's operational requirements.

## 5 Discussion

In this study, we analyzed 45 female health applications (FHAs), encompassing a wide range of health-related functionalities. Our objective was to assess the security and privacy risks associated with FHAs, with a particular focus on permission requests, sensitive data collection, third-party tracking, and privacy policy transparency. In this section, we discuss our findings in relation to the research objectives, highlighting critical concerns that have implications for both user trust and privacy, particularly in the context of female health.

### 5.1 Inconsistencies in Permission Requests

The permission requests in FHAs exhibit significant inconsistencies, with some apps requesting as few as eight permissions, while others request up to 49. Despite offering similar core functionalities, this variance suggests a lack of standardized privacy design across apps. Excessive permission requests, particularly those that are unrelated to the app's stated functionality, expose users to unnecessary privacy risks. Approximately 43% of the detected permissions could not be mapped to the core services offered by these apps, highlighting potential overreach in data collection.

Permissions such as access to location data, audio recordings, media files, and advertising identifiers—especially in apps whose core functionality does not require these permissions—raise concerns about unnecessary data harvesting. This is particularly alarming in the context of female health, where the collection of sensitive health and reproductive data is at stake. The lack of adherence to the principle of data minimization, a core tenant of privacy by design, places users at heightened risk of unauthorized data access.

Additionally, the opacity surrounding permissions that do not directly correlate with core functionality creates a transparency issue. For instance, apps requesting access to location or contact data without clear justification mislead users, diminishing their capacity for informed consent. The disparity in permissions between apps and the overreach in data collection increase the risk of unintended data exposure, creating a larger attack surface for malicious actors. This not only erodes user trust but disempowers individuals, particularly women, who expect higher privacy standards when dealing with reproductive and health-related data.

### 5.2 Over-Collection of Sensitive Data

Our findings indicate that FHAs extensively collect highly sensitive information from users, spanning demographic, reproductive, physical, and mental health categories. Data points include pregnancy tracking, menstrual cycles, mental health conditions, and even sexual activity. Some apps also request information such as race/ethnicity, income, and work history, despite these details being unrelated to the core functionality of the apps. For example, 6 FHAs requested race and ethnicity data, while others requested work or income information, raising significant questions about the necessity and appropriateness of these data points.

Despite the sensitive nature of this data, there is a lack of transparency regarding how it is processed, stored, and shared. Many apps fail to provide clear information on how reproductive and mental health data is handled, leaving users uninformed and vulnerable to potential privacy breaches. The absence of detail surrounding third-party data sharing practices further exacerbates these risks. Additionally, some apps request intimate partner data, such as contraceptive use, without requiring consent from the partner. This creates an ethical dilemma where deeply personal information could be collected without the knowledge or consent of all parties involved, leading to privacy violations.

### 5.3 Risks from Third-Party Trackers

The use of third-party trackers in FHAs presents a severe risk to user privacy. Many of these trackers have a history of data misuse and malicious behavior, posing a substantial threat to users' sensitive information. Our study detected several third-party trackers known for questionable practices, including:

- *InMobi*: Fined by the U.S. Federal Trade Commission for collecting user information without consent[34].
- *IronSource*: Accused of being a malware vector, installing intrusive ads without user knowledge[24].
- *Facebook Ads*: Criticized for enabling manipulation, misinformation, and discrimination[62].
- *RevMob*: Involved in click fraud, inflating advertisement interactions for profit[91].
- *UXCam*: Associated with session replay attacks, recording user interactions for unauthorized purposes[19, 69].

These trackers operate covertly, often collecting and sharing highly sensitive data—such as reproductive and mental health information—with third parties. In most cases, users are not made aware of this data exchange, and apps fail to provide explicit consent mechanisms. The lack of transparency surrounding third-party data sharing is particularly concerning, given that 80% of the apps analyzed did not disclose which entities had access to user data. This hidden data flow increases the risk of exploitation, where intimate details about users' health and reproductive status could be monetized or misused.

### 5.4 Inaccessibility and Ambiguity of Privacy Policies

A fundamental component of privacy is the ability for users to make informed decisions about their data. However, our analysis shows that 20% of FHAs lacked an accessible privacy policy on the Google Play Store, denying users critical information before installation. Furthermore, 45% of apps only provided generic or developer-level privacy policies, which obfuscated the specific data practices for each app. These policies often fail to explain the extent of data collection and sharing, leaving users vulnerable to potential misuse of their sensitive health information.

Even where privacy policies are available, they are often written at a reading level far beyond what is accessible to the average user. This creates a significant barrier, particularly for underserved communities who may already face challenges in accessing healthcare. Moreover, 85% of apps vaguely mentioned data sharing with third parties but did not specify which parties were involved or the purposes behind this sharing. This ambiguity undermines user trust and leaves individuals—especially those in politically sensitive environments—at risk of having their reproductive and health data shared or sold without their knowledge.

## 5.5 Summary and Conclusion

In this study, we investigated the security and privacy risks of FHAs, analyzing permission requests, sensitive data collection, third-party tracking, and privacy policy transparency. Our findings indicate significant privacy risks, including the over-collection of sensitive data, the presence of malicious third-party trackers, and a lack of accessible and transparent privacy policies.

The inconsistencies in permission requests, combined with the overreach in data collection, reflect a lack of adherence to privacy-by-design principles. The pervasive use of third-party trackers, many of which have histories of privacy violations, further exacerbates these concerns. Finally, the inaccessibility and ambiguity of privacy policies prevent users from making informed decisions, undermining user autonomy and trust.

These findings underscore the urgent need for stricter regulatory oversight and ethical design practices within the ecosystem of female health apps. Developers must prioritize transparency, minimize data collection, and ensure that privacy policies are both accessible and clear. Only through such measures can we adequately protect the privacy and security of users, particularly women, who rely on these apps for sensitive health and reproductive information.

## 6 Limitations & Future Work

Our study offers a thorough examination of permissions, third-party trackers, and privacy policies in Female Health Applications (FHAs). However, there are still significant opportunities for future research to build upon these findings in the expanding field of *FemTech* and FHAs. A crucial next step involves regulating FHAs under more stringent health privacy frameworks, such as the Health Insurance Portability and Accountability Act (HIPAA) in the United States. The U.S. government has recently demonstrated its commitment to this goal [39], aiming to ensure that these applications meet the same rigorous standards as other health-related technologies. Furthermore, a comparative study of global legal policies is necessary to investigate how privacy regulations and policies regarding female health data differ across various regions and legal systems. Such an analysis would provide valuable insights into how localized policies shape data practices, especially in jurisdictions with differing levels of digital rights protections.

Our focus on static analysis enabled us to evaluate what permissions and trackers FHAs can *potentially* access, but not necessarily what they *actively* collect or share during actual usage. While static analysis provides valuable insights into potential risks, it does not capture the complete behavior of the applications in dynamic environments. Future work could benefit from conducting dynamic analysis, such as information flow tracking from source to sink, to map how user data is processed, shared, and possibly transmitted to third parties. This type of analysis would provide a more granular understanding of how FHAs handle sensitive user data and whether this behavior aligns with their privacy policies.

Our dataset of 45 applications is notably larger than those used in previous studies of FHAs [7, 27, 59, 80, 98], and we employed a temporal approach to collect data, allowing us to capture how these apps evolve over time. However, our selection criteria, which focused on popular apps in the U.S. market, may limit the generalizability of our findings. We acknowledge that the landscape of FHAs extends beyond the U.S., and there may be regional variations in both app behavior and user privacy concerns. Future research should consider expanding the dataset to include applications from different geographical locations, cultural contexts, and user demographics to provide a more globally representative understanding of the security and privacy risks in this domain.

Despite these limitations, we believe our work establishes a strong foundation for subsequent studies. Our findings offer a valuable frame of reference for further exploration into how FHAs' privacy practices may differ across regions and legal environments. Additionally, researchers can build on this work by investigating the specific ways in which cultural, linguistic, and regulatory differences shape user expectations and app behaviors in diverse settings.

### 6.1 Ethical Considerations

Given the sensitive nature of female health data, ethical considerations were at the forefront of our research. Although we did not involve real users or collect personal data during our analysis, we adhered to ethical research standards throughout the study. No Institutional Review Board (IRB) approval was required, as no human subjects were involved. To conduct our analysis, we used a dedicated test smartphone (Nokia, Android v12 and v13) and created test accounts within the applications. Following the completion of our analysis, all test data and accounts were deleted where possible, and we submitted formal requests to FHAs to ensure the deletion of any remaining data associated with our test accounts.

We are also committed to furthering responsible disclosure. Where our study detected significant privacy concerns, we plan to contact the relevant app developers and inform them of the potential vulnerabilities or overreach in data collection practices. By offering specific recommendations for improving user privacy, we hope to drive positive changes in the industry and contribute to a safer, more transparent ecosystem for female health technologies.

### 7 Related Work

The intersection of women's health technologies and privacy has garnered increasing attention in HCI, particularly in light of recent political and societal shifts regarding reproductive rights. A growing body of research has begun to explore the security and privacy risks posed by FemTech and related applications. For instance, Mehrnezhad et al. [60] examined the risks associated with FemTech, particularly focusing on fertility data and the regulatory frameworks, such as GDPR, that aim to mitigate these risks. Their analysis includes a detailed examination of privacy notices and third-party trackers embedded in fertility applications, highlighting gaps in how privacy is managed.

Similarly, Erickson et al. explored the privacy practices of 11 FemTech applications through privacy policy analysis and network traffic inspection on iPhone devices. Their findings revealed concerning behaviors related to data transmission, raising critical questions about data security and third-party access in FemTech apps.

Our research builds on these foundational studies but significantly extends the scope of inquiry. By analyzing 45 widely-used female health applications, our work moves beyond fertility-specific apps to provide a broader perspective on health apps designed for women (FHAs). Our dataset includes apps covering a wide range of functionalities, from fertility and reproductive health to mental health, fitness, and general physical health management. This expanded focus offers a more holistic view of the security and privacy risks faced by users of FHAs and allows for a comparative analysis across different app categories. Additionally, we explore not only permissions and privacy policies but also embedded third-party trackers, delivering a comprehensive risk assessment.

Several prior studies have focused specifically on the privacy policies of FemTech apps and their implications. Shipp et al. [81] analyzed privacy policies from 30 female health applications and found that reproductive health-related data was often inadequately addressed. Malki et al. [56] examined 20 apps, identifying inconsistencies in how reproductive data was managed and highlighting problematic practices that undermine user privacy. Similarly, Alfawzan et al. [6] reviewed 23 applications and documented poor data-sharing practices, with many apps failing to adhere to basic privacy protection standards.

While policy analysis is essential to understanding the risks and regulatory shortcomings of these technologies, focusing on privacy policies alone offers an incomplete picture. Our study builds on this work by integrating privacy policy analysis with a review of each application's permissions, functionalities, and third-party trackers. This combined approach offers a more comprehensive understanding of the data practices employed by these applications, exposing the broader risks associated with their usage. By situating privacy policies within the context of app behavior—how data is collected, shared, and stored—we provide actionable insights into the design and development of more secure and privacy-conscious FemTech solutions and FHAs in particular.

## 8 Conclusion

In this study, we conducted a comprehensive analysis of 45 popular Female Health Applications (FHAs) for data safety and privacy risks. We evaluated the FHAs based on four dimensions: permissions, trackers, user data collection, and privacy policies. Our findings revealed several alarming issues and challenges in the FHAs domain, such as: excessive and dangerous permissions that could grant the FHAs access to the user's smartphone system and data, and enable them to collect and share sensitive information with trackers and third parties; numerous malicious third-party trackers that pose privacy and security risks for the user data being shared with them; private information that was unrelated to the functionality of the application, and that could violate the user's privacy and expose them to discrimination or harassment; and poorly readable privacy policies that lacked clear and transparent information about data collection and sharing, data retention, user consent, and compliance with regional privacy regulations, and that showed a general lack of adherence to various principles, such as data minimization, purpose limitation, or accountability. To our knowledge, this is the first comprehensive study on overall female health applications, extending the prior work on sub-categories within the domain, such as menstrual cycle or pregnancy apps. We hope our results will help all stakeholders, such as users, developers, regulators, and researchers, improve privacy design by ensuring more informed user policies and data privacy practices.

## A  Policy Coding

This section includes our codes for the privacy policy.

### A.1  Privacy Policy Location and Scope

- **Dedicated-app Policy** = Dedicated privacy policy is available for the applications
- **Multi-app Policy** = Collective privacy policy which is intended for multiple apps including analyzed app
- **Compnay Policy** = A generic policy intended for the developer website and company aiming to cover all products(apps, services, website etc)
- **Generic Policy** = Generic privacy policy which does not discuss the selected application at all

### A.2  Policy regarding 3rd parties

- Information is shared with 3rd parties
- Privacy Policy lists 3rd party libraries and trackers
- Privacy Policy lists explicitly information (email, name, IP, date of birth etc) shared with 3rd party libraries and trackers
- Policy describes functionality and privacy detail about 3rd party.
- Privacy Policy link 3rd party's privacy policy.
- Privacy Policy describe risk factors asssociated with data sharing with 3rd parties

### A.3  Purpose

- Privacy Policy describes purpose and functionality of application.
- Playstore description describe purpose and functionality of the application.*

### A.4 FIPPS Principles

*A.4.1 Access and Rectification.*

- Appropriate access should be provided to PII
- Appropriate opportunities should be provided to correct or amend PII.

*A.4.2 Accountability.*

- Monitor, audit, and document compliance to ensure accountability
- Clear definition of roles and responsibilities with respect to PII for all employees and contractors
- Appropriate training must be provided to all employees and contractors who have access to PII.

*A.4.3 Authority.*

- The authority to handle PII must be identified in the appropriate notice. similar to 2C

*A.4.4 Minimization.*

- Limit the creation, collection, use, processing, storage, maintenance, dissemination, or disclosure of PII to what is necessary to achieve a legally authorized purpose.
- PII should be kept only for as long as needed to achieve its purpose of serving user.

*A.4.5 Quality and Integrity.*

- Accuracy, relevance, timeliness, and completeness of PII should be reasonably ensured
- Ensuring PII quality and integrity is necessary to ensure fairness to the individual

*A.4.6 Individual Participation.*

- Seek individual consent, to the extent practicable, for creating, collecting, using, processing, storing, maintaining, disseminating, or disclosing PII
- Establish procedures for receiving and addressing individuals' privacy-related complaints and inquiries.

*A.4.7 Purpose Specification and Use Limitation.*

- Provide notice of the specific purpose for which each PII is collected
- Explicitly states that PII should only be used, processed, stored, maintained, disseminated, or disclosed for a purpose that is explained in the notice

*A.4.8 Security.*

- Establish administrative, technical, and physical safeguards to protect PII
- Measure put in place to mitigate the harm that can be caused by someone accessing, using, changing, losing, destroying, sharing or exposing the information without authorization.

*A.4.9 Transparency.*

- Provide clear notice about how and why they create, collect, use, process, store, maintain, disseminate, and disclose PII.
- The notice should be easy to understand and easily accessible.

### A.5 Data Subjects Rights

- Privacy policy lists regional privacy regulations (GDPR, CCPA etc)
- Regulations/Laws are from EU/UK (and/or) US
- Application dealing with Children deal with children privacy acts such as COPPA

### A.6 Others

- Age Limit Tag for App on Google Playstore
- Application Privacy Policy restrict app usage to 18 or above.
- Application Privacy Policy restrict app usage to 13 or above.
- Application Privacy Policy does not restrict age.
- Application describe collection, processing and sharing of teenage users data
- Application describe collection, processing and sharing of below 13 users data
- Application notify user or allow them to opt-out when their privacy policy changes

Table 4. Overview of Female Health Apps (FHAs) Dataset.
Labelling: Health & Fitness (☽), Medical (△), and Lifestyle (⊕)

| # | App Name | Developer | # installs | Genre |
|---|----------|-----------|------------|-------|
| 1 | Period Calendar Period Tracker | Simple Design Ltd. | 100,000,000+ | ☽ |
| 2 | Flo Period & Pregnancy Tracker | Flo Health Inc. | 100,000,000+ | ☽ |
| 3 | Workout for Women: Fit at Home | Leap Fitness Group | 50,000,000+ | ☽ |
| 4 | Clue Period Tracker & Calendar | Clue Period Tracker by BioWink | 50,000,000+ | ☽ |
| 5 | Ovulation & Period Tracker | Leap Fitness Group | 10,000,000+ | ☽ |
| 6 | My Calendar - Period Tracker | SimpleInnovation | 10,000,000+ | △ |
| 7 | WomanLog Period Calendar | Pro Active App SIA | 10,000,000+ | ☽ |
| 8 | Ada - check your health | Ada Health | 5,000,000+ | △ |
| 9 | Sweat: Fitness App For Women | Sweat | 5,000,000+ | ☽ |
| 10 | Maya - Period ∣ Pregnancy | Plackal Tech | 5,000,000+ | ☽ |
| 11 | Glow: Track. Shop. Conceive. | Glow Inc | 1,000,000+ | ☽ |
| 12 | Ovulation Tracker App - Premom | premom.com | 1,000,000+ | ☽ |
| 13 | Ovia: Fertility, Cycle, Health | Ovia Health | 1,000,000+ | △ |
| 14 | Workout for Women: Fit & Sweat | Workout Apps | 1,000,000+ | ☽ |
| 15 | Period Tracker & Ovulation | Living Better | 1,000,000+ | △ |
| 16 | Grace Health period tracker | Grace Health | 1,000,000+ | ☽ |
| 17 | Newfemme | New Femme: Empowering Woman | 1,000,000+ | ⊕ |
| 18 | Natural Cycles - Birth Control | NaturalCycles AG | 1,000,000+ | △ |
| 19 | Period Diary Ovulation Tracker | Bellabeat, Inc. | 500,000+ | ☽ |
| 20 | Ovulation & Period Tracker | Magicfit Limited | 100,000+ | ☽ |
| 21 | FEMM Health and Period Tracker | Femm Tech Support | 100,000+ | ☽ |
| 22 | EvolveYou: Strength For Women | EvolveYou App Limited | 100,000+ | ☽ |
| 23 | Health & Her Menopause Tracker | Health & Her | 100,000+ | ☽ |
| 24 | Always You: Period Tracker | Procter & Gamble Productions | 100,000+ | ☽ |
| 25 | Emjoy - Female wellcare | Emjoy | 100,000+ | ☽ |
| 26 | 28 Cycle-Based Wellness | 28 Wellness LLC | 100,000+ | ☽ |
| 27 | Womens Health & Period Tracker | Exandus LLC | 100,000+ | ☽ |
| 28 | Bodylura: Fitness & Nutrition | Body Love Group LLC | 50,000+ | ☽ |
| 29 | Healthshots App | HT Media Ltd - Get Latest & Trending News Updates | 50,000+ | ☽ |
| 30 | StrongHer: Fitness, Yoga, Diet | The Vibe Fitness Limited | 50,000+ | ☽ |
| 31 | Read Your Body | The Body Literacy Collective | 50,000+ | ☽ |
| 32 | FitrWoman | Orreco Limited | 50,000+ | ☽ |
| 33 | MenoLife - Menopause Tracker | MenoLabs LLC. | 50,000+ | ☽ |
| 34 | Silatha: DEI Solutions | Silatha: Training & support to retain your talent | 10,000+ | ☽ |
| 35 | SocialBoat- Women Health Plans | Yoga, Workouts and Diet Plans | 10,000+ | ☽ |
| 36 | How to Have a Healthy Vagina | Zen Jiao | 10,000+ | △ |
| 37 | March: Pelvic & Endometriosis | March Health | 10,000+ | △ |
| 38 | Guava: Health Tracker | Guava Health, Inc. | 10,000+ | ☽ |
| 39 | Hormona - Daily Hormone Health | Hormona | 10,000+ | ☽ |
| 40 | Women's Health Diary 2 | Baskaran Arunasalam | 10,000+ | △ |
| 41 | Orchyd - Track Your Period | Orchyd | 1,000+ | ☽ |
| 42 | Female Fertility Protocols Nat | Dr.Isaac's Holistic Wellness | 10,000+ | ☽ |
| 43 | Nona Woman | Nona Woman | 5,000+ | ☽ |
| 44 | IMC Women's Health | International Medical Center Apps | 5,000+ | ☽ |
| 44 | Bloomth - Wellness & Self-Care | Bloomth | 5,000+ | ☽ |

Fig. 7. Permission which were detected in more than one count of FHA

| App ID | Permissions |
|---|---|
| Common in FHA #2 & #38 | READ_SLEEP, READ_WEIGHT, READ_EXERCISE, READ_HEIGHT, POST_NOTIFICATIONS, READ_RESTING_HEART_RATE, READ_HEART_RATE_VARIABILITY, READ_STEPS, READ_HEART_RATE, READ_BODY_TEMPERATURE, |
| #2 | READ_BASAL_BODY_TEMPERATURE, READ_INTERMENSTRUAL_BLEEDING, WRITE_BASAL_BODY_TEMPERATURE READ_CERVICAL_MUCUS, READ_SEXUAL_ACTIVITY, READ_TOTAL_CALORIES_BURNED, READ_OVULATION_TES WRITE_OVULATION_TEST, READ_MENSTRUATION, READ_ACTIVE_CALORIES_BURNED, WRITE_WEIGHT, WRITE_SEXUAL_ACTIVITY, WRITE_CERVICAL_MUCUS, READ_WHEELCHAIR_PUSHES, WRITE_SLEEP |
| #38 | READ_BLOOD_GLUCOSE, READ_RESPIRATORY_RATE, READ_LEAN_BODY_MASS, READ_OXYGEN_SATURATION, READ_BODY_FAT, READ_BONE_MASS, READ_BLOOD_PRESSURE |

Table 5. Health Permissions

| Category | Information |
|---|---|
| Demographic (43) | Name, Profile Picture, Email, Phone, Location, Country, Gender, Race/Ethnicity, Date of Birth/Age, Employed/Work Info, Relationship Status, Income Info, Financial Hardships, Experience Discrimination, Family Info |
| Reproductive Health (39) | Menstrual Activity, Pregnancy Status, Sexual Activity, Hormone/Harmone Therapy Info |
| Physical Health (41) | Medical History/Health Conditions, Blood Pressure, Weight, Height, Temperature, Diet & Nutrition, Exercise Activity, Wellness Purchase History, Medication, Supplements, Allergies |
| Mental Health (36) | Mental Health Info, Mood Type, Stress Level, Sleep Pattern |
| Lifestyle related Activities (15) | Alcohol & Drugs Activity, Travel Activity |
| Preventive Health Measures (4) | Vaccination History |

Table 6. User-Provided Information Categories and number of FHA for each category

Fig. 8. Permissions and Trackers Across all FHAs

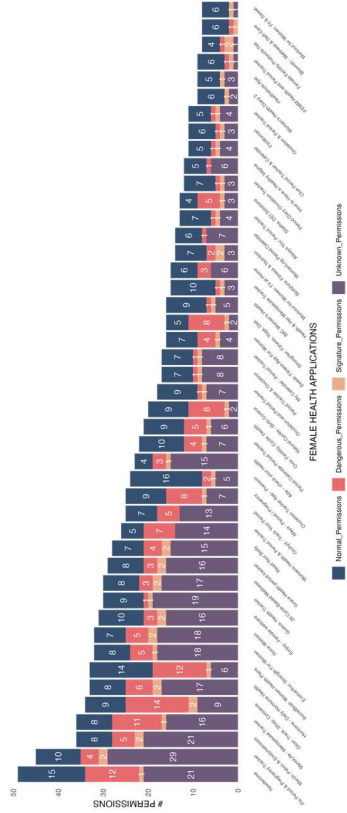

(a) Permissions Across FHAs

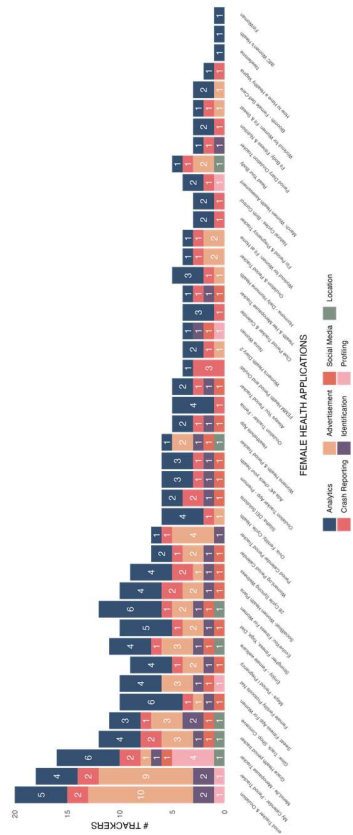

(b) Trackers Across FHAs